\documentstyle[prl,aps,preprint,eqsecnum,rotating,epsfig]{revtex}
\tightenlines

\newcommand{\be}{\begin{equation}}
\newcommand{\ee}{\end{equation}}
\newcommand{\ba}{\begin{eqnarray}}
\newcommand{\ea}{\end{eqnarray}}

\begin{document}
\draft

\title{Cardy-Verlinde Formula and entropy bounds in
 Kerr-Newman-AdS$_4$/dS$_4$ black holes backgrounds}
  \author{Jiliang Jing$^*$ \footnotetext[1]
 {email: jljing@hunnu.edu.cn}}
\address{  Institute of Physics and  Department of Physics,
Hunan Normal University,\\ Changsha, Hunan 410081, P. R. China
 \\ and \\ Department of Astronomy and Applied Physics, University of
Science and Technology of China, \\ Hefei, Anhui 230026, P. R.
China }

\maketitle
\begin{abstract}

The Cardy-Verlinde formula is further verified by using the
Kerr-Newman-AdS$_4$ and Kerr-Newman-dS$_4$ black holes.  In the
Kerr-Newman-AdS$_4$ spacetime, we find that, for strongly coupled
CFTs with AdS duals,  to cast the entropy of the CFT into the
Cardy-Verlinde formula the Casimir energy must contains the terms
$ -n \left({\mathcal{J}} \Omega_H+ \frac{Q\Phi}{2}+
\frac{Q\Phi_0}{2} \right)$, which associate with rotational and
electric potential energies, and the extensive energy includes
the term $-Q \Phi_0$. For the Kerr-Newman-dS$_4$ black hole, we
note that the  Casimir energy is negative but the extensive
energy is positive on the cosmological horizon; while the Casimir
energy is positive but the extensive energy is negative on the
event horizon (the definitions for the two energies possess the
same forms as the corresponding quantities of the
Kerr-Newman-AdS$_4$ black hole). Thus we have to take the absolute
value of the Casimir (extensive) energy in the Cardy-Verlinde
formula for the cosmological (event) horizon. The result for the
Kerr-Newman-dS$_4$ spacetime provides support of the dS/CFT
correspondence. Furthermore, we also obtain the
Bekenstein-Verlinde-like entropy bound for the
Kerr-Newman-AdS$_4$ black hole and the D-bound on the entropy of
matter system in Kerr-Newman-dS$_4$ spacetime. We find that both
the bounds are tightened by the electric charge.

\end{abstract}

\vspace*{0.8cm}

\section{INTRODUCTION}
\label{sec:intro} \vspace*{0.2cm}

Erik Verlinde \cite{Verlinde00} proposed in a recent paper that
the entropy of the conformal field theory (CFT) in a spacetime
 \ba
 ds^2=-dt^2+R^2d\Omega^2_n,
 \ea
can be related to its total energy $E$, Casimir energy $E_c$, and
radius $R$ of the unit sphere $S^n$, which can be explicitly
expressed as
 \ba
 S=\frac{2\pi R}{\sqrt{a_1 b_1}}\sqrt{E_c(2E-E_c)}, \label{scv}
 \ea
where $a_1$ and $b_1$ are two positive coefficients which are
independent of $R$ and $S$. For strong coupled CFT's with AdS
dual, the value of $\sqrt{a_1 b_1}$ is fixed to $n$ exactly. The
expression (\ref{scv}) is referred to as Cardy-Verlinde formula
since it was generalized from the $(1+1)$-dimensional Cardy
formula \cite{Cardy} to the case in arbitrary dimensions.

From the Cardy-Verlinde formula (\ref{scv}) Verlinde  postulated
the Bekenstein-Verlinde entropy bound \cite{Verlinde00}
 \ba
 \frac{S}{2 \pi R E}\leq \frac{1}{\sqrt{a_1 b_1}}. \label{veb}
 \ea
It is shown that CFTs possessing AdS duals satisfy a version of
the Bekenstein entropy bound \cite{Bekenstein}
 \ba
 \frac{S}{2\pi R E} \leq 1. \label{eb1}
 \ea

Witten \cite{Witten98a} pointed out that the energy, temperature,
and entropy of the CFT can be identified with corresponding
quantities of the black hole whose boundary provides a space in
which the CFT resides. Furthermore, conformal invariance can be
invoked to prove that the Bekenstein-Hawking entropy of the black
hole scales with volume of the event horizon. However, we need
additional information to fix the proportional constant between
the entropy and the event horizon volume. It is the
Cardy-Verlinde formula that provides this additional information.

The Cardy-Verlinde formula is vary useful, but it has not been
proved for all CFTs exactly yet. Therefore, the study of validity
of the Cardy-Verlinde formula for every typical spacetimes has
attracted much attention recently \cite{Verlinde00}, \cite{Cai01}
-\cite{Cai01b}: Verlinde \cite{Verlinde00} checked the formula
(\ref{scv}) using the AdS Schwarzschild black holes in various
dimensions and found it holds exactly; Cai \cite{Cai01} studied
the AdS Reissner-Nordstr\"om black holes in arbitrary dimensions
and the AdS black holes in higher derivative gravity, he showed
that the Cardy-Verlinde formula is valid for the  AdS
Reissner-Nordstr\"om black hole if we subtract the electric
potential energy from the Casimir and total energies, but it is
invalid for the AdS black holes in higher derivative gravity;
Birmingham and Mokhtari \cite{Birmingham} found that the
Cardy-Verlinde formula holds for the Taub-Bolt-AdS spacetimes at
high temperature even though these spaces possess some special
thermodynamical properties; Danielsson \cite{Danielsson01}, Cai
\cite{Cai01a}, Medved \cite{Medved01,Medved01a}, Ogushi
\cite{Ogushi01} {\it et al} attempt to generalize the formula to
the case of the de Sitter (dS) black holes. Such studies were
motivated by the observational evidence that our universe has
positive cosmological constant
\cite{Perlmutter,Caldwell,Garnavich} and an interesting proposal
for dS/CFT correspondence \cite{Strominger}; Klemm, Petkou, and
Siopsis \cite{Klemm01} found that the Cardy-Verlinde formula
holds for the one rotation parameter Kerr-AdS$_n$ black holes
whose boundary is a rotating Einstein universe. However, to my
best knowledge, at the moment the question whether or not the
Cardy-Verlinde formula can be used for the Kerr-Newman-AdS$_4$
and the Kerr-Newman-dS$_4$ black holes still remains open. The
main aim of this paper is to settle the question. Another purpose
is to study Bekenstein-Verlinde entropy bound for the
Kerr-Newman-AdS$_4$ black hole and D-bound
\cite{Bousso00a,Bousso00b} for the Kerr-Newman-dS$_4$ black hole.

The paper is organized as follows. In Sec. II, the Cardy-Verlinde
formula and Bekenstein-Verlinde-like entropy bound for the
Kerr-Newman-AdS$_4$ black hole are studied. In Sec. III, the
Cardy-Verlinde formulae on the cosmological and event horizons of
the Kerr-Newman-dS$_4$ black hole are considered, and the D-bound
on the entropy of matter system in the Kerr-Newman-dS$_4$
spacetime is investigated. Some discussions and conclusions are
presented in the last section.

\vspace*{0.4cm}
\section{The Kerr-Newman-AdS$_4$ black hole}
\vspace*{0.5cm}

The Einstein action with a negative cosmological constant along
with the Gibbons-Hawking boundary term on a region of spacetime
$\Sigma$, which with boundary $\partial \Sigma $,  takes the
follows form \cite{Gibbons77}
\begin{equation}
I=\frac{1}{8\pi G_{(d)}}\left( \int_{\Sigma}d^{d}x
\sqrt{-g}\left[ {\mathcal{R}}-F^2-2\Lambda  \right]
-\int_{\partial \Sigma}~d^{d-1}x\sqrt{-\gamma }K \right),
\label{act}
\end{equation}
where $ {\mathcal{R}} $ is the Ricci scalar, $F_{\mu\nu}$ is the
electromagnetic field, $\Lambda $ is the cosmological constant,
which defines a natural length scale as $
\Lambda=-\frac{(d-1)(d-2)}{2\ell^2}$, $\gamma_{ab}$ is the
induced metric on the timelike boundary, and $K$ is the trace of
the extrinsic curvature $K^{ab}$ of the boundary.

The usual charged rotating Kerr-Newman-AdS$_4$ black hole obtained
from the action (\ref{act}) reads in the Boyer-Lindquist
coordinates \cite{Carter,Plebanski}
\begin{eqnarray}
ds^{2} &=&-\frac{\Delta _{r}}{\rho ^{2}}\left(dt-\frac{a}{\Xi
}\sin ^{2}\theta d\phi \right)^{2}+\frac{\rho ^{2}}{\Delta
_{r}}dr^{2}+\frac{\rho ^{2}}{\Delta
_{\theta }}d\theta ^{2} \nonumber  \\
&&+\frac{\Delta _{\theta }\sin ^{2}\theta }{\rho ^{2}}\left[a
dt-\frac{ (r^{2}+a^{2})}{\Xi }d\phi \right]^{2},  \label{kadsmet}
\end{eqnarray}
with
\begin{eqnarray}
\Delta _{r} &=&(r^{2}+a^{2})\left(1+\frac{r^{2}}{\ell^{2}}\right)
-2Mr+q^2,  \nonumber \\
\Delta _{\theta } &=&1-\frac{a^{2}\cos ^{2}\theta}{ \ell^{2}},
  \nonumber \\
\Xi &=&1-\frac{a^{2}}{\ell^{2}},  \nonumber \\
\rho ^{2} &=&r^{2}+a^{2}\cos ^{2}\theta .  \label{delt}
\end{eqnarray}
where the parameters $M$, $a$, and $q$ are related to the mass,
angular momentum, and electric charge parameters of the black
hole, respectively. The event horizon is located at $r_+$, which
is defined as the largest root of the $ \Delta _{r}=0 $. The mass
parameter $M $ in the spacetime (\ref{kadsmet}) and (\ref{delt})
can be written as
\begin{equation}
M=\frac{(r_+^2+a^2)(r_+^2+\ell^2)+q^2 \ell^2}{2r_+
\ell^2}.\label{mass}
\end{equation}
The asymptotic AdS nature of the spacetime
(\ref{kadsmet}) can be exhibited by introducing new coordinates
 \begin{eqnarray}
 & & t= t, \nonumber \\
 & & \Xi y^2\sin^2\Theta=(r^2+a^2)\sin^2 \theta, \nonumber \\
 & & y \cos \Theta=r \cos \theta, \nonumber \\
 & & \Phi=\phi + \frac{a}{\ell^{2}} t.
 \end{eqnarray}

\vspace*{0.2cm} \subsection{Cardy-Verlinde formula}
\vspace*{0.2cm}

Euclideanizing the metric (\ref{kadsmet}) and identifying
$\tau\sim \tau+ \beta_{HK}$ and $\phi\sim
\phi+i\beta_{HK}\Omega_H$, we can get the inverse Hawking
temperature
 \ba
 \beta_{HK}=\frac{4\pi r_+
 \ell^2(r_+^2+a^2)}{3r_+^4+r_+^2(a^2+\ell^2)-(a^2+q^2)\ell^2},
\label{beta}
 \ea
and the angular velocity of the event horizon
 \ba
 \Omega_H=\frac{a\Xi}{r_+^2+a^2}. \label{omg}
 \ea
The Bekenstein-Hawking entropy can be easily obtained form metric
(\ref{kadsmet})
 \ba
 S=\frac{\pi (r_+^2+a^2)}{\Xi }. \label{s1}
 \ea

In order to study Cardy-Verlinde formula, we will use the energy
$E$, angular momentum ${\mathcal{J}}$, and electric charge $Q$,
{\it et al}. The quantities for the Kerr-Newman-AdS$_4$ black
hole can be work out from the action and metric
\cite{Mann00,Das00,Brown93,Brown94,Hawking96,Mann99,balakraus,d=345}.
However, in general the action (\ref{act}) is divergent for the
spacetime. Several kinds of methods for treating the problem were
proposed \cite
{Das00,Brown93,Brown94,Hawking96,Mann99,balakraus,d=345,kls}. Two
attractive solutions to eliminate the divergence in anti-de Sitter
spacetimes are the conformal and counterterm methods\cite{Das00}.
For the Kerr-Newman-AdS$_4$ black hole, the conserved charges,
such as the mass and angular momentum, calculated with either
approach will yield the same values.

Starting from the metric (\ref{kadsmet}), actions (\ref{act}), and
counterterm action $I_{ct}$, Caldarelli, Cognola, and Klemm
\cite{Caldarelli} have computed the mass ${\mathcal{M}}$, the
angular momentum ${\mathcal{J}}_{\phi}$, the electric charge $Q$,
and the electric potential $\Phi$ of the Kerr-Newman-AdS$_4$
black hole, which are given by \cite{Caldarelli}
 \ba
 & &{\mathcal{M}}=\frac{M}{\Xi},\nonumber \\
 & &{\mathcal{J}}_\phi=\frac{M a}{\Xi^2}, \nonumber \\
 & & Q =\frac{q}{\Xi}, \nonumber \\
 & &\Phi_q=\frac{qr_+}{r_+^2+a^2}.
 \label{mjq}
 \ea
We define an electric potential when the rotation goes to be zero
as
 \ba
 \Phi_{q0}=\lim_{a\rightarrow 0}\Phi_q=\frac{q}{r_+}.\label{phi0}
 \ea

We now use above thermodynamical quantities to discuss the
question of whether the entropy of the event horizon for the
Kerr-Newman-AdS$_4$ black hole can be described by the
Cardy-Verlinde formula. According to the AdS/CFT duality
conjecture  \cite{Witten98a,Witten98,Gubser98}, the above
thermodynamical quantities are associated to a strongly coupled
CFT residing on the conformal boundary of the spacetime
(\ref{kadsmet}). The boundary spacetime in which the boundary CFT
resides can be obtained from the bulk metric, up to a conformal
factor. We can rescale the boundary metric so that the finite
volume has a radius $R$. By standard technique we know that the
boundary line element obtained from the metric (\ref{kadsmet})
can be expressed as
 \ba
ds_b^2&=&\lim_{r \rightarrow
\infty}\frac{R^2}{r^2}ds^2=-\frac{R^2}{\ell^2}d t^2 +\frac{2 a
R^2 \sin^2\theta }{\ell^2 \Xi} dt d \phi
+\frac{R^2}{\Delta_\theta}d \theta^2 +\frac{R^2 \sin^2\theta
}{\Xi}d\phi^2, \label{euniverse}
 \ea
which is a rotating Einstein universe.

By the standard technique we know that the temperature $T$, the
energy $E$, angular momentum ${\mathcal{J}}$, the electric
potential $\Phi$, and the non-rotational electric potential
$\Phi_0$ of the CFT must be rescaled by a factor of
$\frac{\ell}{R}$, i.e.,
 \ba
 T=\frac{\ell }{R} T_{HK}, ~~~~ E=\frac{ \ell}{R} {\mathcal{M}},
 ~~~~ {\mathcal{J}}=\frac{ \ell}{R}{\mathcal{J}_\phi}, ~~~~
 \Phi=\frac{\ell }{R} \Phi_q, ~~~~
\Phi_0=\frac{\ell }{R} \Phi_{q0}.
 \label{mjq1}
 \ea
However, the horizon angular velocity and  the entropy are still
given by Eqs. (\ref{omg}) and  (\ref{s1}), respectively.

We define the Casimir energy of the CFT as
 \ba
 E_c=n\left(E+pV-TS-{\mathcal{J}}\Omega_H-\frac{Q\Phi}{2}
 -\frac{Q\Phi_0}{2}\right),\label{ecas}
 \ea
where the pressure is defined as $ p=-\left( \frac{\partial E }
{\partial V } \right)_{S,J,Q} $ and $n=2$. Substituting the
corresponding quantities into Eq. (\ref{ecas}), we get
 \ba
 E_c=\frac{(r_+^2+a^2)\ell }{R \Xi r_+}. \label{ecas01}
 \ea
We also define the extensive energy as
 \ba
E_{ext}=2\left(E-\frac{Q \Phi_0}{2}\right)-E_c, \label{eex}
 \ea
which is in this case
 \ba
 E_{ext}=\frac{(r_+^2+a^2)r_+}{\ell R \Xi }. \label{eex01}
 \ea

In order to further understand the strongly coupled CFTs with AdS
duals, we can write
 \ba
 2\left(E-\frac{Q \Phi_0 }{2}\right)R=\frac{n}{2\pi}\frac{S R}{r_+}\left(1+\frac{1}
 {\Delta^2} \right),\label{eer}
 \ea
and define the ``Casimir entropy" by
 \ba
 S_c=\frac{2\pi}{n}E_c R=S\frac{R}{r_+}.\label{scc}
 \ea
It is obvious that these results possess exactly the behavior of a
two-dimensional CFT \cite{Klemm01} with characteristic scale $R$,
temperature $\tilde{T}=1/(2\pi R \Delta)=r_+/(2\pi r^2)$, and
central charge proportional to $SR/r_+$ (Eq. (\ref{eer}) takes
the same form as Eq. (17) in Ref. \cite{Klemm01}, and the Casimir
energy $S_c$ is essentially proportional to the central charge).

With the Casimir energy (\ref{ecas01}), and the extensive energy
(\ref{eex01}), we find that the entropy of the CFT on the horizon
can be expressed as
 \ba
S&=&\frac{2\pi R}{n}\sqrt{E_c\left[2 \left(E-E_Q\right)
-E_c\right]}\nonumber \\
&=&\frac{\pi (r_+^2+a^2)}{\Xi }, \label{cv}
 \ea
where $E_Q$ is electric potential energy, which is defined as
 \ba
E_Q=\frac{Q\Phi_0}{2}.
 \ea
The difference between expression (\ref{cv}) and the standard
Cardy-Verlinde formula $S=\frac{2\pi R}{n}\sqrt{E_c(2E-E_c)}$ is
that the electric potential energy $E_Q$ emerges in (\ref{cv}),
which must be included in the extensive energy (\ref{eex}) to cast
the entropy of the CFT into the form of the Cardy-Verlinde
formula. We should note that the rotational and electric potential
energies were also contained in the Casimir energy (\ref{ecas})
for case of the Kerr-Newman-AdS$_4$ black hole. We learn that from
Eqs. (\ref{s1}) and (\ref{cv})  the entropy of the CFT agrees
with the Bekenstein-Hawking entropy of the Kerr-Newman-AdS$_4$
black hole. The result provides a conformal field theory
interpretation of the Bekenstein-Hawking entropy. When the
rotation parameter $a$ tends to zero the expressions (\ref{ecas})
and (\ref{cv}) reduce to results of the AdS Reissner-Nordstr\"om
black hole obtained by Cai in Ref. \cite{Cai01}.

\vspace*{0.2cm} \subsection{Bekenstein-Verlinde-like entropy
bound} \vspace*{0.2cm}

Now we study entropy bound of the Kerr-Newman-AdS$_4$ black hole.
As in Ref. \cite{Verlinde00} we can define the
``Bekenstein-Hawking energy" corresponding to the
Bekenstein-Hawking entropy (\ref{s1}) by
 \ba
 E_{BH}=\frac{n S}{2\pi R}= \frac{(r_+^2+a^2)}{ \Xi R}.\label{ebh}
 \ea
From Eqs. (\ref{mass}), (\ref{s1}), (\ref{mjq}),  and  (\ref{ebh})
we find that the energy of the black hole can be expressed as
 \ba
 E&=&\frac{1+\Delta ^2}{2 \Delta} E_{BH}+\frac{Q \Phi_0}{2}
 \nonumber \\
 &=&\frac{1+\Delta ^2}{2 \Delta} E_{BH}+E_Q,
 \ea
where $\Delta=\frac{R}{r_+}$ and $R=\ell$.  We now rewrite the
expression as
 \ba
 E-E_Q=\frac{1+\Delta ^2}{2 \Delta} E_{BH},\label{eeq}
 \ea
and define a entropy $S_B$ which relates to the energy $E-E_Q$ as
 \ba
 S_B&=&\frac{2 \pi}{n} R (E-E_Q). \label{sb}
 \ea
Substituting Eqs. (\ref{ebh}) and (\ref{eeq}) into (\ref{sb}) we
get
 \ba
 S_B &=& \frac{2 \pi}{n} R E_{BH}
 \frac{1+\Delta ^2}{2 \Delta}\nonumber \\
 &=& \frac{1+\Delta ^2}{2 \Delta} S.
 \ea
Furthermore, since we are above the Hawking-Page transition point
\cite{Hawking83}, we have
 \ba
 r_+\geq R,~~~~~\frac{1+\Delta ^2}{2 \Delta}\geq 1 .
 \ea
Thus we obtain
 \ba
 & &E_{BH} \leq (E-E_Q), \nonumber \\
 & &S\leq S_B=\frac{2\pi}{n} R \left(E-\frac{q^2}{2\Xi r_+}\right),
 \label{bound}
 \ea
where the equality holds when the Hawking-Page phase transition
is reached.   When $q=0$ and $a=0$ the entropy bound
(\ref{bound}) reproduces precisely the Bekenstein-Verlinde one
(\ref{veb}) for the nonrotating neutral object. It is obvious that
the bound is tightened by the electric charge $q$.

\vspace*{0.4cm}
\section{The Kerr-Newman-dS$_4$ black hole}
\vspace*{0.5cm}

Strominger \cite{Strominger} proposed the dS/AdS correspondence
which says that there is a dual between quantum gravity on a dS
space and a Euclidean CFT on a boundary of the dS space. Although
much effort was spent recently, we know that the dS/CFT
correspondence so far acquired quite incomplete. Thus, the study
in this section only provide support of the dS/CFT
correspondence.

The metric of the Kerr-Newman-dS$_4$ black hole can be obtained
from the (\ref{kadsmet}) by replacing $\ell^2$ with $-\ell^2$,
which can be written as
\begin{eqnarray}
ds^{2} &=&-\frac{\Delta _{r}}{\rho ^{2}}\left(dt-\frac{a}{\Xi
}\sin ^{2}\theta d\phi \right)^{2}+\frac{\rho ^{2}}{\Delta
_{r}}dr^{2}+\frac{\rho ^{2}}{\Delta
_{\theta }}d\theta ^{2} \nonumber  \\
&&+\frac{\Delta _{\theta }\sin ^{2}\theta }{\rho ^{2}}\left[a
dt-\frac{ (r^{2}+a^{2})}{\Xi }d\phi \right]^{2},  \label{kdsmet}
\end{eqnarray}
with
\begin{eqnarray}
\Delta _{r} &=&(r^{2}+a^{2})\left(1-\frac{r^{2}}{\ell^{2}}\right)
-2Mr+q^2,  \nonumber \\
\Delta _{\theta } &=&1+\frac{a^{2}\cos ^{2}\theta}{ \ell^{2}},
  \nonumber \\
\Xi &=&1+\frac{a^{2}}{\ell^{2}},  \nonumber \\
\rho ^{2} &=&r^{2}+a^{2}\cos ^{2}\theta .  \label{kdelt}
\end{eqnarray}
where the parameters $M$, $a$, $q$ have the same means as the
case of the Kerr-Newman-AdS$_4$ black hole. The mass parameter $M
$ in the spacetime (\ref{kdsmet}) can be expressed as
\begin{equation}
M=-\frac{(r_+^2+a^2)(r_+^2-\ell^2)-q^2 \ell^2}{2r_+
\ell^2}.\label{kmass}
\end{equation}

\vspace*{0.2cm} \subsection{Cardy-Verlinde formula on cosmological
horizon} \vspace*{0.2cm}

The thermodynamical quantities associated the cosmological horizon
are
 \ba
 & &\beta_{c}=\frac{4\pi r_c
\ell^2(r_c^2+a^2)}{3r_c^4+r_c^2(a^2-\ell^2)+(a^2+q^2)\ell^2},
\nonumber \\
 & & {\mathcal{M}}=-\frac{M}{\Xi}, \nonumber  \\
 & &\Omega_c=-\frac{a\Xi}{r_c^2+a^2}, \nonumber  \\
 & & {\mathcal{J}}_\phi=\frac{M a}{\Xi^2}, \nonumber \\
 & & Q =\frac{q}{\Xi}, \nonumber  \\
 & & \Phi_q=-\frac{q r_c}{r_c^2+a^2}, \nonumber  \\
 & & \Phi_{q0}=-\lim_{a\rightarrow 0}\Phi_q=-\frac{q}{r_c},
 \label{ctherm}
 \ea
while the entropy is
 \ba
 S=\frac{\pi (r_c^2+a^2)}{\Xi }. \label{cks1}
 \ea

The temperature $T$, the energy $E$, angular momentum
${\mathcal{J}}$, the electric potential $\Phi$, and the
non-rotational electric potential $\Phi_0$ of the CFT must be
rescaled by a factor of $\frac{\ell}{R}$ as.

Substituting Eqs. (\ref{ctherm}) and (\ref{cks1}) into  the
Casimir energy
$E_c=n(E+pV-TS-{\cal{J}}\Omega_c-Q\Phi/2-Q\Phi_0/2)$, we get
 \ba
 E_c=-\frac{(r_c^2+a^2)\ell }{R \Xi r_c}, \label{ckecas01}
 \ea
while the extensive energy $2(E-Q\Phi_0/2)-E_c$ becomes
 \ba
 E_{ext}=\frac{(r_c^2+a^2)r_c}{\ell R \Xi }. \label{ckeex01}
 \ea
Thus it is easy to see that the entropy of the CFT on the
cosmological horizon is given by
 \ba
S=\frac{2\pi R}{n}\sqrt{|E_c|[2 \left(E-E_Q\right) -E_c]},
\label{ckcv}
 \ea
The only one difference between Eqs. (\ref{cv}) and (\ref{ckcv})
is that we take the absolute value of the Casimir energy in Eq.
(\ref{ckcv}).

\vspace*{0.2cm} \subsection{Cardy-Verlinde formula on black hole
horizon} \vspace*{0.2cm}

The thermodynamical quantities associated the black hole horizon
are given by
 \ba
 & &\beta_{HK}=-\frac{4\pi r_+
\ell^2(r_+^2+a^2)}{3r_+^4+r_+^2(a^2-\ell^2)+(a^2+q^2)\ell^2},
\nonumber \\
 & &{\mathcal{M}}=\frac{M}{\Xi},\nonumber  \\
 & & \Omega_H=\frac{a\Xi}{r_+^2+a^2}, \nonumber  \\
 & &{\mathcal{J}}_\phi=\frac{M a}{\Xi^2}, \nonumber \\
 & & Q =\frac{q}{\Xi}, \nonumber  \\
 & &\Phi_q=\frac{qr_+}{r_+^2+a^2}, \nonumber  \\
 & &\Phi_{q0}=\lim_{a\rightarrow 0}\Phi_q=\frac{q}{r_+}.\label{therm}
 \ea
The entropy is
 \ba
 S=\frac{\pi (r_+^2+a^2)}{\Xi }. \label{ks1}
 \ea

The temperature $T$, the energy $E$, angular momentum
${\mathcal{J}}$, the electric potential $\Phi$, and the
non-rotational electric potential $\Phi_0$ of the CFT must be
rescaled by a factor of $\frac{\ell}{R}$ as we did in the case of
the Kerr-Newman-AdS$_4$ black hole.

Substituting the thermodynamical quantities listed in Eqs.
(\ref{therm}) and (\ref{ks1}) into definition of the Casimir
energy
 \ba
 E_c=n\left(E+pV-TS-{\mathcal{J}}
\Omega_H-\frac{Q\Phi}{2} -\frac{Q\Phi_0}{2}\right),
 \ea
we get
 \ba
 E_c=\frac{(r_+^2+a^2)\ell }{R \Xi r_+}, \label{kecas01}
 \ea
while the extensive energy
 \ba
 E_{ext}=2\left(E-\frac{Q
\Phi_0}{2}\right)-E_c,
 \ea
takes the value
 \ba
 E_{ext}=-\frac{(r_+^2+a^2)r_+}{\ell R \Xi }. \label{keex01}
 \ea
We note that the extensive energy is negative now. With the
Casimir energy (\ref{kecas01}) and the extensive energy
(\ref{keex01}), the entropy of the CFT on the event horizon can
be cast to
 \ba
S&=&\frac{2\pi R}{n}\sqrt{E_c|2 \left(E-E_Q\right)
-E_c|}\nonumber \\
 &=& \frac{\pi (r_+^2+a^2)}{\Xi }, \label{kcv}
 \ea
where $E_Q=\frac{Q\Phi_0}{2}$. The only one difference between
the Cardy-Verlinde formulae (\ref{cv}) and (\ref{kcv}) is that we
take the absolute value of the extensive energy in Eq.
(\ref{kcv}).

\vspace*{0.2cm} \subsection{D-bound of entropy} \vspace*{0.2cm}

By applying the classical Geroch process to the cosmological
horizon, Bousso \cite{Bousso00a,Bousso00b} proposed the D-bound on
the entropy of matter system in the Schwarzschild-de Sitter space.
The D-bound shows that the entropy of objects in dS space is
bounded by the difference of the entropies in the exact dS space
and in the asymptotically dS space
 \ba
 S_m\leq S_0-S_c, \label{csb}
 \ea
where $S_0$ is the entropy of the exact dS space and $S_c$ is the
cosmological horizon entropy when the matter is present.

We now apply the D-bound to the Kerr-Newman-dS$_4$ black hole.
The cosmological horizon radius is determined by the maximal root
of the function
 \ba
 (1+\frac{a^{2}}{r_c^{2}})\left(1-\frac{r_c^{2}}{r_0^{2}}\right)
-\frac{2M}{r_c}+\frac{q^2}{r_c^2}=0 ,\label{kebh}
 \ea
where $r_0\equiv \sqrt{\ell^2}$. The equation can be rewritten as
 \ba
 \frac{r_0^2}{r_c^2+a^2}&=&\left(1-\frac{2M}{r_c}+\frac{q^2+a^2}
 {r_c^2}\right)^{-1}\nonumber \\
& \approx & 1
 +\frac{2M}{r_c}-\frac{q^2+a^2}{r_c^2}, \label{con}
 \ea
where we use the large cosmological horizon limit:
 \ba
 \frac{2M}{r_c}\ll 1,~~~~~~\frac{q^2}{r_c^2}\ll 1,
 ~~~~~~~\frac{a^2}{r_c^2}\ll 1.\label{con1}
 \ea
We learn from Eqs. (\ref{con}) and (\ref{con1}) that
 \ba
 r_0^2-\frac{r_c^2+a^2}{\Xi}\leq 2 r_c \left[\frac{M}{\Xi}
 -\frac{q^2}{2r_c \Xi}\right]. \label{con2}
 \ea
By using Eqs. (\ref{csb}) and (\ref{con2}), we obtain an entropy
bound of the rotating charged object in the Kerr-Newman-dS$_4$
space: \ba
 S\leq S_B&=&2 \pi R \left(\frac{M}{\Xi}
 -\frac{q^2}{2R \Xi}\right), \label{dsbound}
 \ea
where we replace $r_c$ by $R$.  When the rotational parameter
$a=0$ the bound (\ref{dsbound}) reduces to the D-bound for the
Reissner-Norstr\"om de-Sitter space \cite{Cai01ab}; and when
$a=0$ and $q=0$ it reproduces precisely the Bekenstein entropy
bound (\ref{eb1}).

\section{conclusion and discussion}

Starting from the four dimensional Kerr-Newman-AdS and
Kerr-Newman-dS black holes, we further verify the Cardy-Verlinde
formula which relates the entropy of the conformal field theory
to its Casimir energy $E_c$, the energy $E$ (or the extensive
energy), and the radius $R$ of the unit sphere $S^n$. For
strongly coupled CFTs with AdS duals, we find that in order to
apply the Cardy-Verlinde formula to case of the
Kerr-Newman-AdS$_4$ black hole, the Casimir energy (\ref{ecas})
must contains the terms $ -n \left({\mathcal{J}} \Omega_H+
\frac{Q\Phi}{2}+\frac{Q\Phi_0}{2}\right)$, which relates to
rotational and electric potential energies of the CFT. The
difference between expression (\ref{cv}) and the standard
Cardy-Verlinde formula (\ref{scv}) is that the non-rotational
electric potential energy $E_Q=\frac{Q \Phi_0}{2} $ emerges in
the formula (\ref{cv}). When the rotation parameter vanishes the
expressions (\ref{ecas}) and (\ref{cv}) reduce to results of the
AdS Reissner-Nordstr\"om black hole obtained in Ref. \cite{Cai01}.

For the Kerr-Newman-dS$_4$ black hole, the definitions of the
Casimir and the extensive energies possess the same form as that
of the Kerr-Newman-AdS$_4$ black hole. However, we note that the
Casimir energy (\ref{ckecas01}) is negative but the extensive
energy (\ref{ckeex01}) is positive on the cosmological horizon;
while the Casimir energy (\ref{kecas01}) is positive but the
extensive energy (\ref{keex01}) is negative on the event horizon.
Thus we must take the absolute value of the Casimir energy in the
Cardy-Verlinde formula (\ref{ckcv}) and absolute value of the
extensive energy in the Cardy-Verlinde formula (\ref{kcv}).  As
the rotation parameter $a$ goes to zero the results reduce to
that of the Reissner-Nordstr\"om-dS black hole found by Cai in
Ref.\cite{Cai01b}. The results provide support of the dS/AdS
correspondence.

The fact that the entropy of the CFT agrees precisely with the
Bekenstein-Hawking entropy also provides a conformal field theory
interpretation of the Bekenstein-Hawking entropies of the
Kerr-Newman-AdS$_4$ and Kerr-Newman-dS$_4$ black holes. The
result supports the conclusion of the reference \cite{Jing},
which says that for the Kerr-Newman-AdS and Kerr-Newman-dS black
holes the statistical entropies obtained from the density of
states determined by conformal field theory method agree with
their Bekenstein-Hawking entropies.

We also study the entropy bounds of these spacetimes. The
Bekenstein-Verlinde-like entropy bound (\ref{bound}) for the
Kerr-Newman-AdS$_4$ black hole is obtained by comparing different
kinds of the energies. It is obvious that the bound is tightened
by the electric charge. For the non-rotating neutral object the
entropy bound (\ref{bound}) reduces to the Bekenstein-Verlinde one
\cite{Verlinde00}. On the other hand, the D-bound on the entropy
of matter system in Kerr-Newman-dS$_4$ space is also found, which
is described by formula (\ref{dsbound}). The bound is also
tightened by the charge. For the non-rotating neutral object the
bound (\ref{dsbound}) reproduces precisely the Bekenstein entropy
bound.

\vspace*{0.4cm}

\begin{acknowledgements}
This work was supported by the National Natural Science Foundation
of China under Grant No. 19975018, and Theoretical Physics Special
Foundation of China under Grant No. 19947004.
\end{acknowledgements}

\end{document}